\documentclass[aps,prl,twocolumn]{revtex4-1}
\usepackage{blindtext}

\usepackage{graphicx}
\usepackage{amssymb}
\usepackage{amsmath}
\usepackage{bm}
\usepackage{braket}
\usepackage{gensymb}
\usepackage{footmisc}
\usepackage{booktabs}
\usepackage{array}
\usepackage{hyperref}
\usepackage{xcolor}
\hypersetup{
    colorlinks,
    linkcolor={black},
    citecolor={blue},
    urlcolor={blue}
}

\begin{document}

\title{Band Gap Closing in a Synthetic Hall Tube of Neutral Fermions}

\author{Jeong Ho Han, Jin Hyoun Kang, and Y. Shin}
\email{yishin@snu.ac.kr}

\affiliation{
Department of Physics and Astronomy and Institute of Applied Physics, Seoul National University, Seoul 08826, Korea
\\
Center for Correlated Electron Systems, Institute for Basic Science, Seoul 08826, Korea
}


\begin{abstract}
We report the experimental realization of a synthetic three-leg Hall tube with ultracold fermionic atoms in a one-dimensional optical lattice. The legs of the synthetic tube are composed of three hyperfine spin states of the atoms, and the cyclic inter-leg links are generated by two-photon Raman transitions between the spin states, resulting in a uniform gauge flux $\phi$ penetrating each side plaquette of the tube. Using quench dynamics, we investigate the band structure of the Hall tube system for a commensurate flux $\phi=2\pi/3$. Momentum-resolved analysis of the quench dynamics reveals that a critical point of band gap closing as one of the inter-leg coupling strengths is varied, which is consistent with a topological phase transition predicted for the Hall tube system.
\end{abstract}

\maketitle
Ultracold atoms in optical lattices have become a unique platform for studying condensed matter physics in a clean and controllable environment~\cite{Jaksch2005,Bloch2008}. Over the past decade, many experimental techniques have been demonstrated to generate artificial gauge potentials for neutral atoms, providing an interesting opportunity for exploring topologically nontrivial states of matter~\cite{Dalibard2011}. The Hofstadter-Harper (HH) Hamiltonian, which is the essential model for quantum Hall physics, was realized in two-dimensional (2D) optical lattice systems using laser-assisted tunneling~\cite{Aidelsburger2013,Miyake2013,Lin2011,Goldman2014}. Recently, ladder systems with the HH Hamiltonian, dubbed Hall ribbons, were demonstrated in the synthetic dimension framework~\cite{Boada2012,Celi2014}; in this framework, the internal degrees of freedom of atoms such as hyperfine spins~\cite{Mancini2015,Stuhl2015} and clock states~\cite{Livi2016} are exploited as a virtual lattice dimension and the hopping along the dimension is provided by laser-induced couplings between the internal states. The framework was further extended with the external degrees of freedom of atoms such as momentum states~\cite{Meier2016,An2017-1,An2017-2} and lattice orbitals~\cite{Kang2018}.

The key advantage of using synthetic lattice dimensions is versatile boundary manipulation. Sharp edges can be defined and individually detected with state-sensitive imaging, thus allowing for experimental investigation of various phenomena such as chiral edge currents~\cite{Mancini2015,Stuhl2015}, topological solitons at interfaces~\cite{Meier2016}, and magnetic reflection~\cite{An2017-1,An2017-2}. Furthermore, nontrivial lattice geometries can be created in synthetic dimensions, which are hardly achievable with conventional optical lattices but may give rise to novel topological states~\cite{Boada2015,Barbarino2018}. A remarkable example is a ladder geometry with a periodic boundary condition (PBC), which can be realized by cyclically connecting the synthetic lattice sites. It is under a PBC that a Hall lattice system exhibits a true fractal structure of the single-particle energy spectrum, called Hofstadter's butterfly~\cite{Hofstadter1976}. Additionally, Laughlin's pump, which is an ideal manifestation of quantized Hall conductivity and corresponding Chern number, has been proposed for a torus geometry~\cite{Laughlin1981,Zeng2015,Taddia2017}.

In this paper, we report the experimental realization of a synthetic Hall lattice system of a tube geometry with ultracold fermionic atoms. In our scheme, the neutral fermions are confined in a one-dimensional (1D) optical lattice and three hyperfine spin states are employed as a synthetic dimension to form a three-leg tube structure. The cyclic links between the legs are created by spin-momentum couplings via two-photon Raman transitions between the spin states, and a uniform gauge flux $\phi=2\pi/3$ per side plaquette is generated, thus realizing an HH Hamiltonian with a PBC~\cite{Hofstadter1976}. Using quench dynamics, we investigate the band structure of the synthetic Hall tube system. When the system evolves from a symmetric tube to an open ladder as one of the inter-leg coupling strengths is decreased, we observe a critical point of band gap closing, which is consistent with a topological phase transition predicted for the Hall tube system. This work opens a new avenue for studies of topological phases with ultracold atoms in unconventional lattice geometries.

Our experiment starts with preparing a degenerate Fermi gas of $^{173}$Yb atoms in the $\left|F=5/2,m_F=-5/2\right>$ hyperfine spin state of the $^1\mathrm{S}_0$ ground energy level~\cite{Lee2017}. The typical atom number is $N\approx1.0\times10^4$ and the temperature is $T/T_\mathrm{F}\approx0.3$, where $T_\mathrm{F}$ is the Fermi temperature of the trapped sample. The atoms are adiabatically loaded in a three-dimensional optical lattice potential generated by superposing three orthogonal standing waves with periodicity $d_{x,z}=\lambda_\mathrm{L}/2$ and $d_y=\lambda_\mathrm{L}/\sqrt{3}$, where $\lambda_\mathrm{L}=532~$nm is the laser wavelength. The final lattice depths are $(V_x,V_y,V_z)=(5,20,20)E_{\mathrm{L},\alpha}$, where $E_{\mathrm{L},\alpha}=h^2/(8md^2_\alpha)$ for $\alpha\in\left\{x,y,z\right\}$, $h$ is the Planck constant, and $m$ is the atomic mass. Because tunneling along the $y$ and $z$ directions is highly suppressed by large lattice depths, our lattice system is effectively 1D. The tunneling amplitude is $t_x=2\pi\times264~$Hz, and the characteristic filling factor is estimated to be $\approx 0.75$ with trapping frequencies of $(\omega_x,\omega_y,\omega_z)=2\pi\times(58,42,132)$~Hz~\cite{Kohl2005,Bloch2008}. An external magnetic field of 153~G is applied along $\hat{z}$ to lift the spin degeneracy of the $^1$S$_0$ energy level.

\begin{figure}
\includegraphics[width=8.2cm]{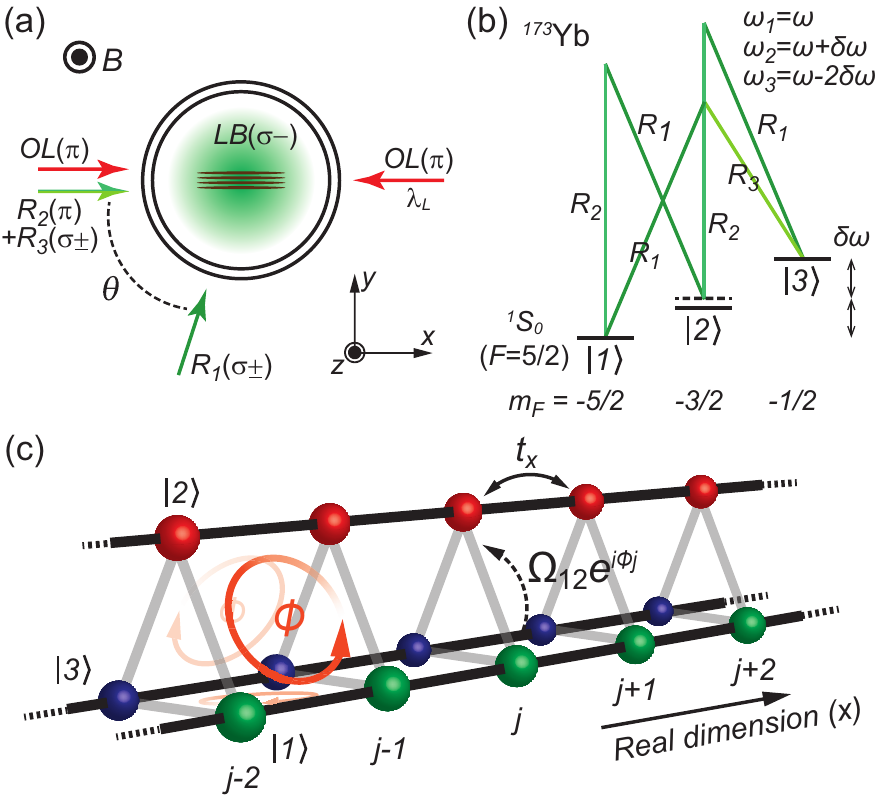}
\caption{Realization of a synthetic Hall tube with neutral atoms. (a) Schematic of the experimental setup. Fermionic $^{173}$Yb atoms are confined in an optical lattice and illuminated by three Raman laser beams $R_{1,2,3}$. A magnetic field $B$ and an additional laser light (LB) are applied along $\hat{z}$ to control the energy levels of the spin states. (b) The three lowest spin states of $^{173}$Yb are coupled to each other via two-photon Raman transitions by $R_{1,2,3}$. (c) Synthetic three-leg Hall tube with a uniform gauge flux $\phi$ on each side plaquette. The three legs are formed by the three spin states of the atoms in the 1D optical lattice (black lines) and the inter-leg tunneling with complex amplitude (gray lines) is provided by the cyclic Raman couplings between the spin states. The flux $\phi$ is controlled by the Raman beam angle $\theta$ in (a) (see the text for details).\label{fig:scheme}}
\end{figure}

The three lowest spin states, which we denote $\left|1\right>\equiv\left|m_F=-5/2\right>$, $\left|2\right>\equiv\left|m_F=-3/2\right>$, and $\left|3\right>\equiv\left|m_F=-1/2\right>$, are employed for the three legs of the synthetic tube system. To generate inter-leg couplings, three linearly polarized Raman laser beams \textit{R}$_{1,2,3}$ are irradiated on the sample~[Fig.~\ref{fig:scheme}(a)], where the wave vectors of the laser beams are given by $\vec{k}_{r1}=k_R(\cos \theta \hat{x} +\sin\theta \hat{y})$ and $\vec{k}_{r2}=\vec{k}_{r3}=k_R\hat{x}$, respectively, and the polarization directions are horizontal for \textit{R}$_{1,3}$ and vertical for \textit{R}$_{2}$ to the $xy$ plane. The laser frequencies of $R_{1,2,3}$ are set to $\omega_1=\omega$, $\omega_2=\omega+\delta\omega$, and $\omega_3=\omega-2\delta\omega$, respectively, where $\omega$ is the laser frequency blue-detuned by $1.97$~GHz from the $\left|^1\mathrm{S}_0,F=5/2\right>\rightarrow\left|^3\mathrm{P}_1,F'=7/2\right>$ transition line. When $\delta\omega$ is tuned to half of the energy difference between $\left|1\right>$ and $\left|3\right>$, the three spin states $\left\{\left|1\right>,\left|2\right>,\left|3\right>\right\}$ can be resonantly coupled to each other in a cyclic manner by two-photon Raman transitions, as described in Fig.~\ref{fig:scheme}(b). Thus, a three-leg synthetic tube is constructed with the fermions in the 1D optical lattice~[Fig.~1(c)].

In the synthetic tube system, the Raman coupling between the spin states $|s\rangle$ and $|s'\rangle$ is described by inter-leg tunneling with complex amplitude $\Omega_{ss'}e^{i\phi j}$, where $\Omega_{ss'}$ is the Rabi frequency of the corresponding two-photon Raman transition and $j$ is the site index for the real lattice. The spatial phase modulations of the tunneling amplitude originate from the momentum transfer $\hbar \Delta \vec{k}$ of the two-photon transition, yielding $\phi = (\Delta\vec{k}\cdot\hat{x})d_x$~\cite{Goldman2014}. In our experimental setup, $\Delta\vec{k}=\vec{k}_{r2,r3}-\vec{k}_{r1}=k_R[(1-\cos\theta)\hat{x}-\sin\theta\hat{y}]$ for all the cyclic inter-leg couplings and $\phi=2\pi k_R d_x (1-\cos\theta)$ regardless of spin state. When a fermion hops around any side plaquette of the tube, it acquires a uniform net phase of $\phi$, thus realizing the HH Hamiltonian in the tube geometry. In this work, we set the Raman beam angle $\theta\approx72.6^\circ$ to have $\phi=2\pi/3$, satisfying the PBC for the synthetic dimension. Because the $\sigma$-$\sigma$ transition ($\Delta m_F$=$2$) for the $\left|1\right>$-$\left|3\right>$ coupling is relatively weak, the intensity ratio of $R_{1,2,3}$ is adjusted to create a symmetric coupling structure. We measure $\Omega_{12}=\Omega_{31}\approx12.3 t_x$. Here, $\Omega_{23}/\Omega_{12}$ is fixed because the $\pi$-$\sigma$ ($\Delta m_F$=$1$) transitions for the $|1\rangle$--$|2\rangle$ and $|2\rangle$--$|3\rangle$ couplings are created by the same pair of Raman beams, and the ratio is nearly unity within 1.2\%.

In realizing the three-leg Hall tube, careful control of the energy levels of the spin states is necessary to suppress the optical transitions to the other spin states, $\left|4\right>\equiv \left|m_F=1/2\right>$ and $\left|5\right>\equiv \left|m_F=3/2\right>$. The energy level $\nu_s$ of spin state $\left|s\right>$ is determined by the sum of the magnetic Zeeman shift and the total AC Stark shift due to laser radiation. To generate sufficiently large differential AC Stark shifts, we apply to the sample an additional laser radiation along $\hat{z}$~\cite{Song2018}, which is $\sigma^-$--polarized and detuned by $-70~$MHz with respect to the $\left|^1\mathrm{S}_0,F=5/2\right>\rightarrow\left|^3\mathrm{P}_1,F'=7/2\right>$ transition line. Under the final experimental condition, the energy level differences between the spin states are spectroscopically measured~\cite{supp} and $\left(\xi_{1},\xi_{2},\xi_{3},\xi_4,\xi_5\right)$$\approx$$\left(0,-0.2,0,-2, 1.7\right)\Omega_{12}$, where $\xi_s = \left(\nu_s - \nu_1\right)-(s-1)\delta \omega$ and $\delta\omega=2\pi \times 30.4$~kHz. $\xi_s$ is the detuning of $|s\rangle$ from the energy staircase formed by two-photon Raman processes with a step unit of $\delta \omega$. The atom loss rate into $|4\rangle$ and $|5\rangle$ is measured to be $\approx 0.01 \Omega_{12}$.

\begin{figure}
\includegraphics[width=8.5cm]{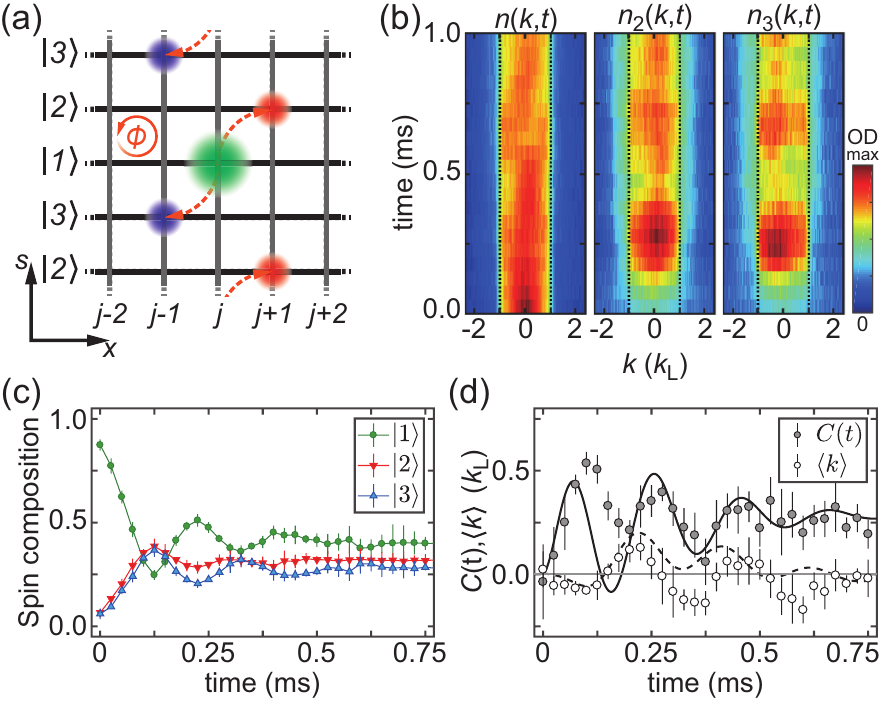}
\caption{Quench dynamics of the three-leg Hall tube for $\phi=2\pi/3$. (a) Illustration of the atomic motion in the Hall tube. Atoms are initially prepared in the spin-$\left|1\right>$ leg and the inter-leg couplings are suddenly activated. (b) Time evolution of the lattice momentum distribution $n(k,t)$ of the sample, $n_2(k,t)$ of the atoms in $|2\rangle$, and $n_3(k,t)$ of the atoms in $\left|3\right>$. Time evolution of (c) the fractional spin populations, (d) the average lattice momentum $\left<k\right>$ of the sample, and the difference $\mathcal{C}(t)=\langle k_2\rangle -\langle k_3\rangle$ between the momenta of the two legs $|2\rangle$ and $|3\rangle$. Each data point comprises five measurements of the same experiment, and the error bar is their standard deviation. The solid and dashed lines in (d) show the numerical simulation results for $\mathcal{C}$ and $\langle k \rangle$, respectively, including phenomenological damping~\cite{supp}.\label{fig:cyclic}}
\end{figure}

\begin{figure}
\includegraphics[width=8.5cm]{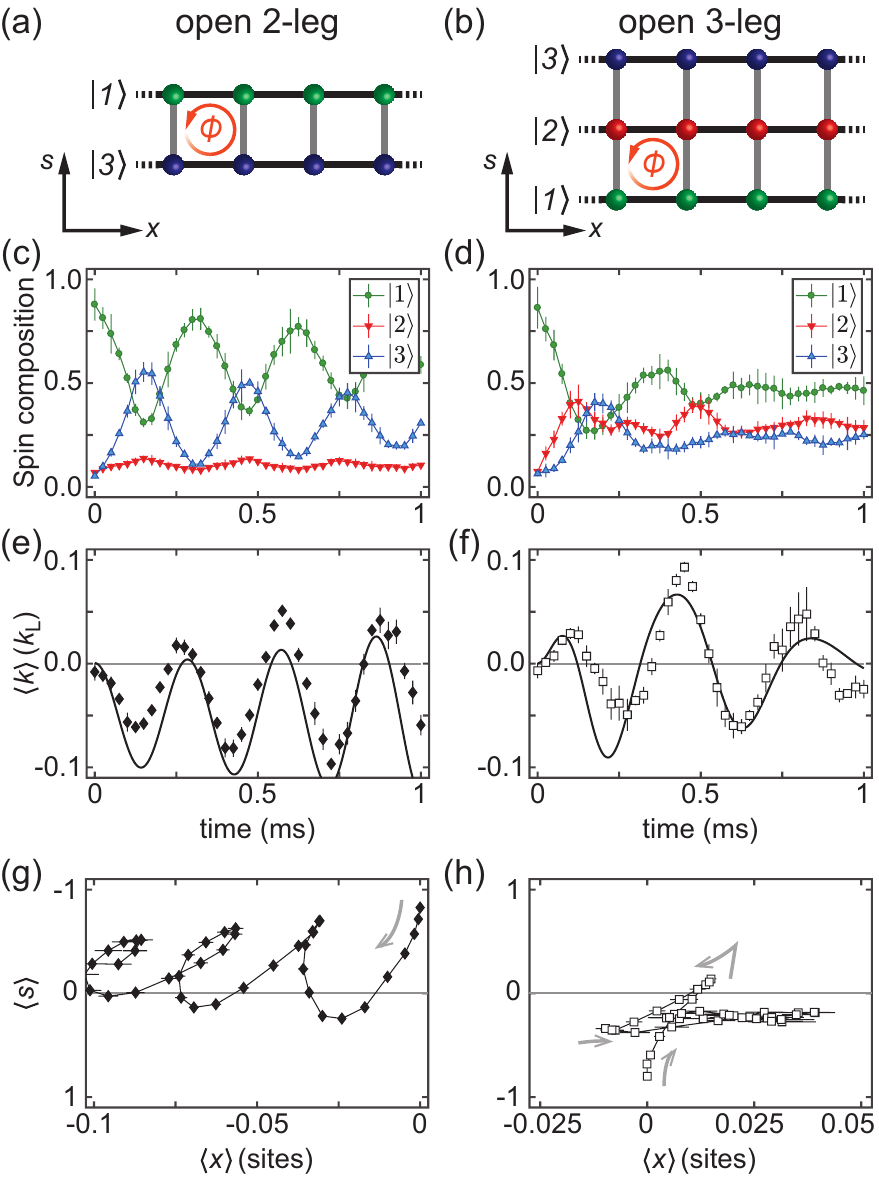}
\caption{Quench dynamics of (a) two-leg and (b) three-leg ladders with open boundaries for $\phi=2\pi/3$. Time evolution of (c,d) the fraction spin populations and (e,f) the average lattice momentum $\left<k\right>$. The solid lines display the numerical simulation results for $\langle k \rangle$~\cite{supp}. Each data point comprises five measurements of the same experiment. (g,h) Trajectories of the ladder systems in the plane of the spin and real lattice positions $\langle s \rangle$ and $\langle x \rangle$. $\left<s\right>=\tilde{n}_3-\tilde{n}_1$, where $\tilde{n}_s$ is the fractional population of spin component $|s\rangle$ and $\left<x\right>$ is calculated from $\left<k\right>$ using the knowledge of band dispersion~\cite{Mancini2015}.\label{fig:open}}
\end{figure}

The Bloch Hamiltonian of the three-leg Hall tube system is given by
\begin{equation} \label{eq:Hamiltonian}
\begin{aligned}
&\hat{H}_q/\hbar = \\
&\begin{pmatrix}
-2t_x\cos{\left(q-\phi\right)} & \Omega_{12}/2 & \Omega_{31}/2 \\ 
\Omega_{12}/2 & \xi_2-2t_x\cos{\left(q\right)}  & \Omega_{23}/2\\ 
\Omega_{31}/2 & \Omega_{23}/2 & -2t_x\cos{\left(q+\phi\right)}
\end{pmatrix}
\end{aligned},
\end{equation}
where $q$ is the quasimomentum of the lattice tube system normalized by $d_x^{-1}$~\cite{supp}. For a symmetric case with $\Omega_{12}=\Omega_{23}=\Omega_{31}$ and $\xi_2=0$, it is well known that the Hamiltonian $\hat{H}_q$ for $\phi=2\pi/3$ embeds a topologically nontrivial state, which is protected by a generalized inversion symmetry~\cite{Barbarino2018,Nourse2016}. In our experiment, this symmetry is preserved with spatially uniform $\Omega_{ss'}$ and the topological state survives $\xi_2\neq 0$, featuring a nonzero Zak phase $Z=1$ of its lowest band~\cite{Zak1989}. In $\hat{H}_q$, time-reversal symmetry, particle-hole symmetry and chiral symmetry are broken, which corresponds to the symmetry class A (unitary) of the Altland-Zirnbauer classification~\cite{Altland1997,Chiu2016}. When the lowest band is completely filled, the system represents a topologically insulating state analogous to the integer quantum Hall state~\cite{Schnyder2008}.

To demonstrate the presence of a gauge flux on plaquettes, we investigate the quench dynamics of the synthetic Hall system. Atoms are initially prepared in the leg $\left|1\right>$, and then the inter-leg couplings are suddenly activated by turning on the Raman laser beams. After a variable hold time, the spin composition of the sample is measured by imaging with optical Stern-Gerlach spin separation~\cite{Taie2010}, and separately, the lattice momentum distribution $n(k)$ of the sample is measured using an adiabatic band-mapping technique~\cite{Kohl2005,supp}. Note that in the band-mapping process, the quasimomentum state with $q$ is transformed into a superposition of free-space momentum states of the three spin states in the first Brillouin zone (BZ), where the momentum $k_s$ of spin state $|s\rangle$ is related to $q$ as $k_s d_x=[q+(s-2)\phi]$ modulo $2\pi$ and $-k_\mathrm{L}<k_s\leq k_\mathrm{L}$ with $k_\mathrm{L}=\pi/d_x$. The momentum distribution $n_s(k)$ of the atoms in $\left|s\right>$ is also measured by spin-selective imaging~[Fig.~\ref{fig:cyclic}(b)]~\cite{Lee2017}.

The measurement results of the time evolution of the quenched synthetic Hall tube system are displayed in Figs.~\ref{fig:cyclic}(c) and \ref{fig:cyclic}(d). At the early time $t<100~\mu$s, when the atoms start transferring to the legs $\left|2\right>$ and $\left|3\right>$, the average lattice momentum of the sample, $\left<k\right>=\int^{k_\mathrm{L}}_{-k_\mathrm{L}} k n(k)dk /\int^{k_\mathrm{L}}_{-k_\mathrm{L}} n(k) dk$, shows no significant variations; however, the difference between the momenta of the atoms transferred into $\left|2\right>$ and $\left|3\right>$, $\mathcal{C}(t) = \left<k_2\right> - \left<k_3\right>$, where $\left<k_s\right>=\int^{k_\mathrm{L}}_{-k_\mathrm{L}} k n_s(k)dk /\int^{k_\mathrm{L}}_{-k_\mathrm{L}} n_s(k) dk$, increases noticeably. This means that the atoms in the legs $\left|2\right>$ and $\left|3\right>$ move in positive and negative directions of the real lattice, respectively, which is understandable based on the classical motion of a charged particle moving in the tube in the presence of a magnetic field~[Fig.~\ref{fig:cyclic}(a)]. At later times, the spin composition and $\mathcal{C}(t)$ show damped oscillations, which are reasonably accounted for by a numerical simulation for $\hat{H}_q$ including phenomenological damping~\cite{supp}. The asymmetry between $|2\rangle$ and $|3\rangle$ and the small oscillations of $\left<k\right>$ result from nonzero $\xi_2$.

The quench evolution of the Hall system is further examined for open ladder geometries [Figs.~\ref{fig:open}(a) and \ref{fig:open}(b)]. The structure modification is achieved by deactivating two or one of the inter-leg links; by shifting $\omega_2$ ($\omega_3$) by $2\pi \times 400$ ($-400$)~kHz, a two-(three-)leg  ladder is formed. For large detuning, the associated inter-leg couplings are effectively turned off but the AC Stark shifts due to the Raman beams are nearly unaffected~\cite{Wang2012,Cheuk2012}. The time evolutions of the spin composition and the average momentum $\left<k\right>$ are displayed in Figs.~\ref{fig:open}(c)--\ref{fig:open}(f). In contrast to the Hall tube case, $\left<k\right>$ shows relatively large oscillations because the atoms are initially prepared at an edge of the ladder. Interestingly, $\left<k\right>$ changes its sign during the oscillations, and the behavior is well captured by the numerical simulations [Fig.~\ref{fig:open}(e) and \ref{fig:open}(f)]. In the open three-leg ladder case, we attribute the behavior mainly to the large gauge flux $\phi>\pi/2$ causing atoms to reflect at the BZ boundary. We note that the sign change of $\langle k \rangle$ was not observed in a previous experiment for a smaller gauge flux~\cite{Mancini2015}. In Figs.~\ref{fig:open}(g) and \ref{fig:open}(h), the semiclassical trajectories of the ladder systems are displayed in the plane of spin and real lattice positions. The open two-leg ladder case shows damped cyclotron motion truncated by the ladder edge, and the three-leg case exhibits bouncing motions due to the Bloch oscillations in the course of cyclotron motion. These observations corroborate the presence of a gauge flux on the side plaquettes of the synthetic tube.

\begin{figure}
\includegraphics[width=8.5cm]{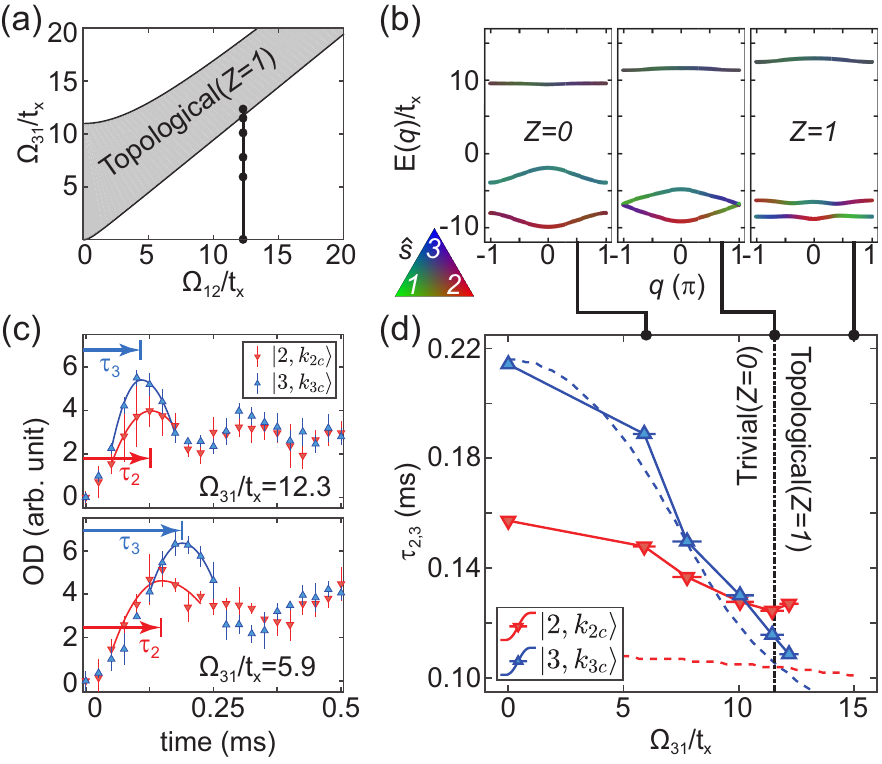}
\caption{Observation of band gap closing in the three-leg Hall tube. (a) Phase diagram of the system. The shaded area indicates a topologically nontrivial phase with a nonzero Zak phase $Z=1$ of the lowest band. The solid dots indicate the parameter positions explored in the experiment. (b) Band structures calculated for various $\Omega_{31}$ with $\Omega_{12}=\Omega_{23}\approx12.3 t_x$. A topological phase transition occurs together with band gap closing at $q_c=\pm\pi$. (c) Quench evolution of $n_2(k_{2c})$ and $n_3(k_{3c})$, where $k_{2c}$ and $k_{3c}$ are the lattice momenta of $|2\rangle$ and $|3\rangle$, respectively, corresponding to $q_c$. Each data point is obtained by averaging five measurements of the same experiment and the error bar is their standard deviation. The time $\tau_{2}$ ($\tau_3$) for the first maximum of $n_2(k_{2c})$ ($n_3(k_{3c})$) is determined by fitting the experimental data to an asymmetric parabola function (solid line). (d) $\tau_{2}$ and $\tau_3$ as functions of $\Omega_{31}$. The red and blue dashed lines show the numerical results for $\tau_2$ and $\tau_3$, respectively. The crossing of $\tau_{2}$ and $\tau_{3}$ reveals the critical point of band gap closing.\label{fig:transition}}
\end{figure}

In Fig.~\ref{fig:transition}(a), we present the phase diagram of the Hall tube system for $\phi=2\pi/3$ in the plane of $\Omega_{12}$ and $\Omega_{31}$. The topological phase with $Z=1$ exists in a region of $\Omega_-<\Omega_{31}<\Omega_+$, where the boundaries are given by $\Omega_\pm=\pm3t_x -\xi_2 + \sqrt{(3t_x \mp \xi_2)^2 + \Omega^2_{12}}$. Our current system with $\Omega_{31}=\Omega_{12}\approx12.3 t_x$ is located in the topological regime and its transition to a topologically trivial phase with $Z=0$ can be driven by, for example, decreasing $\Omega_{31}$ below the critical value of $\Omega_-=11.6 t_x$~\cite{Barbarino2018}. In Fig.~\ref{fig:transition}(b), the band dispersions of the Hall tube system are displayed for various $\Omega_{31}$, showing that the topological phase transition at $\Omega_{31}=\Omega_-$ occurs with closing the energy gap between the first and second bands at quasimomentum $q_c=\pm\pi$~\cite{Footnote2}. According to the bulk-edge correspondence, band gap closing is a generic and necessary feature of the topological phase transition of a symmetry-preserving system~\cite{Ezawa2013}.

The critical point of band gap closing is probed via momentum-resolved analysis of the quench dynamics. As the band gap closes, the dynamic evolution for $q=q_c$ is governed by a single energy scale that is determined by the energy difference between the third band and the two touching lowest bands. Therefore, the gap closing would be characteristically reflected in the quench evolution of the spin composition at $q_c$. The momenta of the spin states $\left|2\right>$ and $\left|3\right>$ corresponding to $q_c=\pm\pi$ are $k_{2c}=-k_\mathrm{L}$ and $k_{3c}=-k_\mathrm{L}/3$, respectively, and we measure the quench evolution of $n_{2}(k_{2c})$ and $n_{3}(k_{3c})$ for various $\Omega_{31}\leq \Omega_{12}$~[Fig.~\ref{fig:transition}(c)]. When $\Omega_{31}$ is decreased by decreasing the intensity of $R_3$, the resulting reduction of the AC Stark shift is compensated for by applying another off-resonant laser light with the same polarization as $R_3$. To obtain the characteristic time scales of the spin composition oscillations, we determine the times $\tau_{2}$ and $\tau_{3}$ at which $n_2(k_{2c})$ and $n_3(k_{3c})$ reach their first maxima, respectively, by fitting the experimental data to an asymmetric parabolic function~\cite{Footnote3}.

Figure~\ref{fig:transition}(d) shows the measurement results of the time scales as functions of $\Omega_{31}$. At $\Omega_{31}=\Omega_{12}$, $\tau_3$ is smaller than $\tau_2$ and increases faster than $\tau_2$ as $\Omega_{31}$ decreases. The crossing of $\tau_2$ and $\tau_3$ occurs at $\Omega_{31}\approx 10.4 t_x$ in the vicinity of the expected critical point $\Omega_-$. The numerical simulation reproduces the observed crossing behavior of the two time scales and yields $\tau_2=\tau_3$ at $\Omega_{31}=\Omega_-$, which validates our experimental approach using the time scales of quench dynamics to probe band gap closing. The deviation of the measured critical value from the predicted $\Omega_-$ is not clearly understood. This might be due to imperfection in spin-selective imaging, the damping, or the interaction effects in the quench dynamics, which are neglected in our numerical simulations. We note that for our experimental parameters, the on-site interaction energy is estimated to be $U/\hbar\approx1.7t_x$.

In conclusion, we realize a synthetic three-leg Hall tube with $\phi=2\pi/3$ and demonstrate the band gap closing at a critical point of the topological phase transition of the system. In our experimental setup, the gauge flux $\phi$ can be controlled by $\theta$, and we expect an immediate expansion of this work to study fractal band structures with varying magnetic fluxes from commensurate to incommensurate values. Further studies may include interatomic interactions~\cite{Barbarino2015}, which are expected to show fractional charge behavior~\cite{Zeng2015}, using the recently implemented orbital Feschbach resonance~\cite{Pagano2015,Hofer2015}.

We thank Moosong Lee for early contributions to this work and Seji Kang for experimental assistance. This work is supported by the Institute for Basic Science (IBS-R009-D1) and the National Research Foundation of Korea (Grant Nos. NRF-2018R1A2B3003373, 2014-H1A8A1021987).

\bibliographystyle{apsrev4-1}

\begin{references}
\bibitem{Footnote2} For $\Omega_{12}=\Omega_{23}$, the $\hat{H}_q$ is mirror symmetric along spin dimension with respect to $\left|2\right>$, and therefore the quasimomentum for gap closing should be $q=0$ or $\pm\pi$. Here, the topological transition at $\Omega_{31}=\Omega_+$ shows the gap closing at $q=0$.
\bibitem{Footnote3} The fitting function is given by $f(t) = \alpha_1 (t-\tau_s)^2 + \beta$ for $t\leq t_s$ and $\alpha_2 (t-\tau_s)^2 + \beta$ for $t>t_s$ with four fitting parameters, $\alpha_{1,2}$, $\beta$, and $\tau_s$.
\end{references}

\newpage


\begin{center}
\textbf{\large Supplemental Material}
\end{center}

\setcounter{equation}{0}
\setcounter{figure}{0}
\setcounter{table}{0}
\makeatletter
\renewcommand{\theequation}{S\arabic{equation}}
\renewcommand{\thefigure}{S\arabic{figure}}

\subsection*{Experimental sequence}

\begin{figure}
\includegraphics[width=7.6cm]{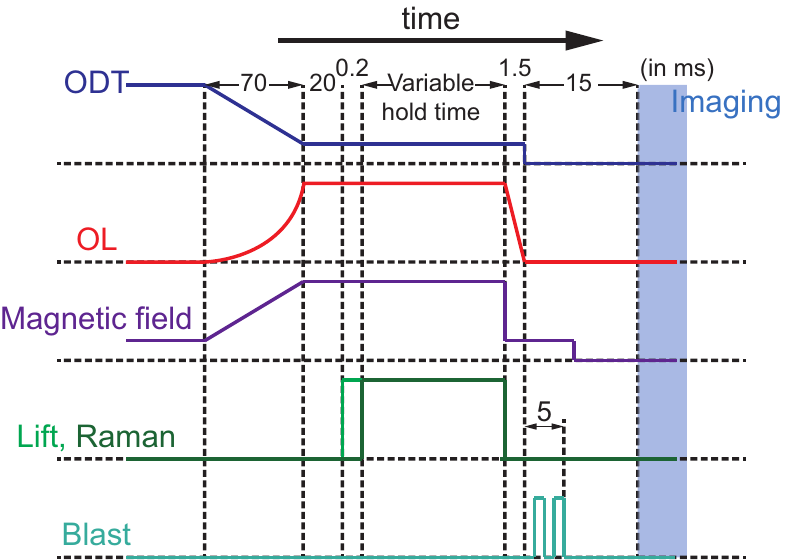}
\caption{ Schematic of the experimental sequence. The figure is not to scale.\label{fig:seq}}
\end{figure}

A schematic of the experimental sequence is presented in Fig.~\ref{fig:seq}. First, we prepare a degenerate Fermi gas of $^{173}$Yb in the $|1\rangle \equiv |F=5/2, m_F=-5/2\rangle$ hyperfine ground state in a crossed optical dipole trap (ODT) using forced evaporative cooling and optical pumping techniques~\cite{Lee2017}. The total atom number is $N\approx 1.0\times10^4$, and the sample temperature is $T/T_\mathrm{F}\approx0.3$, where $T_\mathrm{F}$ is the Fermi temperature of the trapped sample. The fractional population of the atoms in the other spin states with $m_F\neq -5/2$ is less than 3\%. We adiabatically load the atomic cloud in a 3D orthorhombic optical lattice (OL) with lattice constants of $d_x=d_z=\lambda_\mathrm{L}/2$ and $d_y=\lambda_\mathrm{L}/\sqrt{3}$, where $\lambda_\mathrm{L}=532~$nm is the laser wavelength. The lattice potential is exponentially ramped up in $70$~ms to the target depth $\left(V_x,V_y,V_z\right)$=$\left(5,20,20\right)E_\mathrm{L,\alpha}$, where $E_\mathrm{L,\alpha}=h^2/8md^2_{\alpha}$ is the lattice recoil energy for the $\alpha\in\left\{x,y,z\right\}$ direction. The adiabaticity of the lattice loading is confirmed by the fact that the sample temperature is not significantly altered even after reversing the loading sequence. $V_\alpha$ is calibrated by a modulation spectroscopy method~\cite{Heinze2011}. During the lattice ramp-up, we reduce the ODT depth to counteract the increase in the overall trapping potential due to the OL and apply an external magnetic field of $153$~G along $\hat{z}$ to lift the spin degeneracy of the $^1$S$_0$ ground level, resulting in a Zeeman energy splitting of $h\times31.6~$kHz between adjacent spin states. We hold the atoms in the final lattice potential for another $20~$ms to ensure equilibrium. At this stage, the sample is in a metallic state with a characteristic filling factor of $N \prod_\alpha \frac{d_\alpha}{ \zeta_\alpha} \approx 0.75$, where $\zeta_{\alpha}=\sqrt{2\hbar t_{\alpha}/(m\omega_{\alpha}^2)}$, $t_\alpha$ is the tunneling amplitude, and $\omega_\alpha$ is the trapping frequency of the harmonic trapping potential.

Next, we turn on the $\sigma^-$--polarized laser beam, which is referred to as a Lift beam (LB), and after $200~\mu$s, we switch on the Raman laser beams. The role of the Lift beam is to generate differential AC stark shifts of the spin states, which is necessary for suppressing unwanted Raman transitions, in particular, to the $\left|4\right>\equiv\left|m_F=1/2\right>$ spin state. The Lift laser beam is detuned by $-70$~MHz from the $\left|^1\mathrm{S}_0,F=5/2\right>\rightarrow\left|^3\mathrm{P}_1,F'=7/2\right>$ transition line and its intensity is 8.5~mW/cm$^2$. In the quench experiment of the three-leg Hall tube, the fractional spin populations of $\left|4\right>$ and $\left|5\right>\equiv|m_F=3/2\rangle$ are measured to be less than 13\% and 7\%, respectively, after 1~ms evolution. The beam waists of the Lift and Raman beams are $150~\mu$m, much larger than the \textit{in situ} sample radius of $15~\mu$m; thus, the confining effect due to the inhomogeneous intensity distributions of the laser beams is negligible. Under the Lift beam, the lifetime of the atoms in the optical lattice is measured to be $\approx 250$~ms, which is approximately four times shorter than that of the atoms without the Lift beam. As the atoms are illuminated by the Raman laser beams, their lifetime is further reduced to $\approx 20$~ms in the open three-leg ladder case and even down to $\approx 6$~ms in the three-leg Hall tube case.

The lattice momentum distribution of the system is measured employing a conventional band-mapping technique~\cite{Kohl2005}. Following the sudden turn-off of the Lift and Raman beams, the lattice potential is linearly ramped to zero within 1.5~ms, and an absorption image is obtained after a time-of-flight of 15~ms using the $^1$S$_0$$\rightarrow$$^1$P$_1$ transition. For spin-selective imaging, the atoms not in the target spin state are removed by applying short pulses of laser light resonant with the $\left|^1\mathrm{S}_0,F=5/2\right>\rightarrow\left|^3\mathrm{P}_1,F'=7/2\right>$ transition within the initial $5$~ms of the free expansion. The removal process causes  inter-spin collisions, which results in atom position blurring in the absorption image [Fig.~2(a)].

\subsection*{Measurement of energy level differences}

\begin{figure}
\includegraphics[width=7cm]{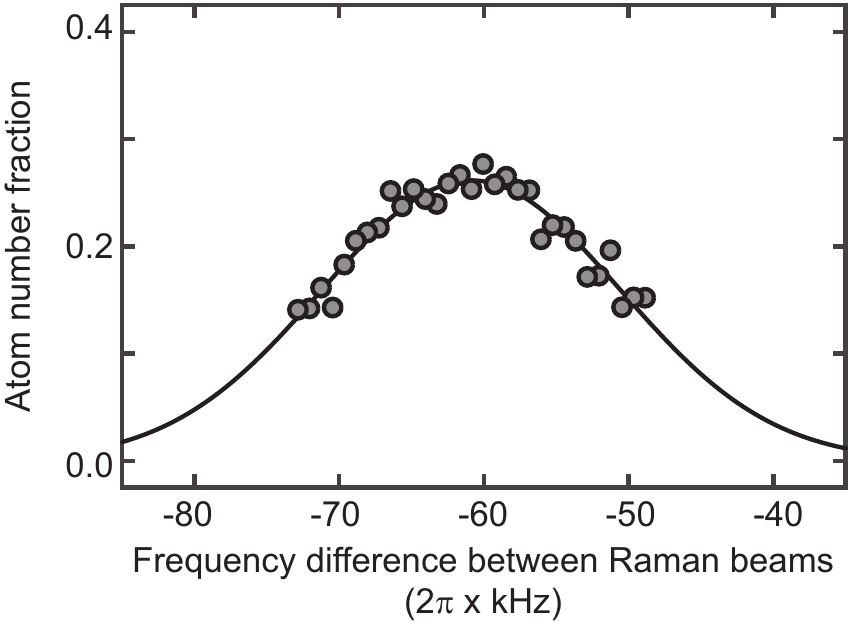}
\caption{ Raman spectrum of a spin-polarized sample in $\left|1\right>\equiv\left|m_F = -5/2\right>$. The fractional population of the atoms in $\left|3\right>\equiv\left|m_F=-1/2\right>$ is plotted as a function of the frequency difference of the Raman laser beams. The Raman beam pulse duration is $t_0=50~\mu$s. The solid line indicates a Gaussian curve fit to the data. \label{fig:levels}}
\end{figure}

We measure the energy level differences between the spin states by two-photon Raman spectroscopy. For a spin-polarized atomic cloud prepared in the presence of the Lift beam but without the optical lattice, we apply a short pulse of the Raman laser beams with a pulse duration of $t_0=50~\mu$s and we measure the fractional population of the atoms transferred to the target spin state using an optical Stern-Gerlach spin separation method as a function of the frequency difference $\delta \omega_r$ of the two Raman laser beams associated with the transition. Here, the frequency of the other Raman laser beam which is not involved in the target transition is set to be far detuned to prevent Raman transitions to other spin states, while its AC Stark shift effect is maintained. Figure~\ref{fig:levels} shows a typical Raman spectrum for the $\left|1\right> \rightarrow \left|3\right>$ transition, where $|3\rangle \equiv |m_F=-1/2\rangle$. The center frequency $\delta \omega_{r,c}$ is determined by fitting a Gaussian function to the spectrum, and taking into account the kinetic energy contribution, the energy level difference $\nu_3-\nu_1$ between the two spin states is obtained as $\nu_3-\nu_1 = \delta \omega_{r,c} - \frac{\hbar}{2m}[2 k_R \sin (\theta/2)]^2$, where $k_R$ is the wave number of the Raman beams and $\theta$ is the angle between the two Raman beams. In determining the energy level $\nu_4$ ($\nu_5$) of the spin state $|4\rangle$ ($|5\rangle$), we use a spin-polarized atomic sample in $|3\rangle$. For our experimental condition, we measure $\left(\xi_{1},\xi_{2},\xi_{3},\xi_4,\xi_5\right)$$\approx$$\left(0,-0.2,0,-2, 1.7\right)\Omega_{12}$, where $\xi_s=\left(\nu_s - \nu_1\right)-(s-1)\delta \omega$, $\delta\omega\equiv\nu_3/2=2\pi\times30.4~$kHz, and $\Omega_{12}=2\pi\times3.3~$kHz. $\Omega_{12}$ is the Rabi frequency of the $\left|1\right>$--$\left|2\right>$ Raman coupling ($|2\rangle\equiv |m_F=-3/2\rangle$).

\subsection*{Tight-binding model}
In a rotating wave approximation, the tight-binding model Hamiltonian for our synthetic three-leg Hall tube system is given by 
\begin{equation} \label{eq:tb}
\begin{split}
\hat{H}/\hbar&=\sum_{j}\sum_{s=1}^3 \left(-t_x  \hat{c}^\dagger_{j+1,s}\hat{c}_{j,s} + \mathrm{h.c.}\right)\\
&+\sum_{j} \sum_{s=1}^2 \left(\frac{\Omega_{s,(s+1)}}{2}e^{i\phi j}\hat{c}^\dagger_{j,s+1}\hat{c}_{j,s}+ \mathrm{h.c.}\right)\\
&+\sum_{j} \left( \frac{\Omega_{31}}{2}e^{i\phi j}\hat{c}^\dagger_{j,1}\hat{c}_{j,3}+ \mathrm{h.c.}\right)\\
&+\sum_{j}\sum_{s=1}^3 \left(\xi_s + \epsilon_j/\hbar\right) \hat{c}^\dagger_{j,s} \hat{c}_{j,s} + \frac{U}{2\hbar}\sum_{j}\sum_{s \neq s'} \hat{n}_{j,s}\hat{n}_{j,s'},
\end{split}
\end{equation}
where $\hat{c}_{j,s}$ ($\hat{c}^\dagger_{j,s}$) is the annihilation (creation) operator for a fermion in the Wannier state $\left|j,s\right>$ localized at the real lattice site $j=1,...,L_x$ with spin $s=1,2,3$. The first term represents tunneling in the real lattice; the second and third terms describe the inter-leg couplings generated by the Raman laser beams, where $\Omega_{ss'}$ is the Rabi frequency of the two-photon Raman transition between the spin states $\left|s\right>$ and $\left|s'\right>$ and the position-dependent complex phase factor $e^{i \phi j}$ results from the momentum imparted by the Raman transition; the fourth term is the on-site energy in the rotating frame, including the external trapping potential contribution, $\epsilon_j$; and the last term is the on-site interaction energy with number operator $\hat{n}_{j,s}\equiv\hat{c}^\dagger_{j,s}\hat{c}_{j,s}$.

Under a unitary transformation $\hat{\mathcal{U}} \hat{c}_{j,s} \hat{\mathcal{U}}^\dagger$=$e^{i\phi (s-2) j}\hat{c}^\prime_{j,s}$, the Hamiltonian is re-expressed as
\begin{equation} \label{eq:tb2}
\begin{split}
\hat{H}^\prime/\hbar&=\sum_{j}\sum_{s=1}^3 \left(-t_x e^{-i\phi(s-2)}  \hat{c}^{\prime\dagger}_{j+1,s}\hat{c}^\prime_{j,s} + \mathrm{h.c.}\right)\\
&+\sum_{j} \left(\frac{\Omega_{12}}{2}\hat{c}^{\prime\dagger}_{j,2}\hat{c}^\prime_{j,1}+ \frac{\Omega_{23}}{2}\hat{c}^{\prime\dagger}_{j,3}\hat{c}^\prime_{j,2} + \mathrm{h.c.}\right)\\
&+\sum_{j} \left( \frac{\Omega_{31}}{2}e^{i3\phi j}\hat{c}^{\prime\dagger}_{j,1}\hat{c}^\prime_{j,3}+ \mathrm{h.c.}\right)\\
&+\sum_{j}\sum_{s=1}^3 \xi_s \hat{c}^{\prime\dagger}_{j,s} \hat{c}^\prime_{j,s},
\end{split}
\end{equation}
where the external potential and interaction terms are neglected. When $\phi=2\pi/3$, the complex phase factor $e^{i3\phi j}$ in the third term becomes unity and $j$-independent, and via a transformation $\hat{c}^\prime_{q,s}$=$\frac{1}{\sqrt{L_x}}\sum_{j}e^{iqj}\hat{c}^\prime_{j,s}$, $\hat{H}^\prime$ can be represented in momentum space by the 3-by-3 Bloch Hamiltonian
\begin{equation} \label{eq:HamiltonianMainText}
\begin{aligned}
&\hat{H}_q/\hbar = \\
&\begin{pmatrix}
\xi_1-2t_x\cos{\left(q-\phi\right)} & \Omega_{12}/2 & \Omega_{31}/2 \\ 
\Omega_{12}/2 & \xi_2-2t_x\cos{\left(q\right)}  & \Omega_{23}/2\\ 
\Omega_{31}/2 & \Omega_{23}/2 & \xi_3-2t_x\cos{\left(q+\phi\right)}
\end{pmatrix}
\end{aligned}.
\end{equation}

\begin{figure}[t]
\includegraphics[width=8.4cm]{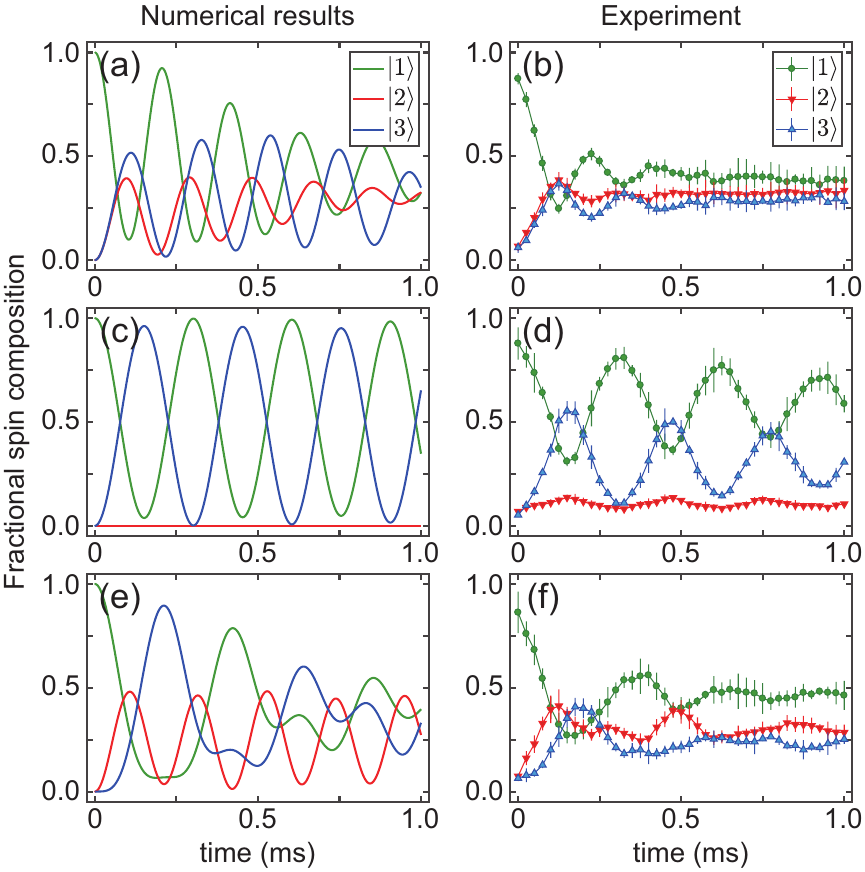}
\caption{ Calculated quench evolution of the fractional spin composition for various boundary conditions: (a) three-leg Hall tube, (c) open two-leg ladder, and (e) open three-leg ladder. (b, d, f) Corresponding experimental data shown in Fig.~2(c), Fig.~3(c) and 3(d), respectively.\label{fig:simulation_spin}}
\end{figure}

\subsection*{Numerical simulation}
We perform a numerical simulation of the quench dynamics by solving the Bloch equation,
\begin{equation} \label{eq:Blocheq}
i\hbar\frac{\partial}{\partial t}\begin{pmatrix}
c_1(q,t)\\ 
c_2(q,t)\\ 
c_3(q,t)
\end{pmatrix} = \hat{H}_q \begin{pmatrix}
c_1(q,t)\\ 
c_2(q,t)\\ 
c_3(q,t)
\end{pmatrix}.
\end{equation}
The atomic density $n_s(k_s,t)$ for spin $s$ and momentum $k_s$ is calculated as $n_s(k_s)=|c_s(q)|^2$, where $k_s d_x=[q+(s-2)\phi]$ modulo $2\pi$ and $-k_\textrm{L} < k_s \leq k_\textrm{L}$ with $k_\textrm{L}=\pi/d_x$. The initial conditions for $c_s$ at $t=0$ are set as $c_1(q,0)=\sqrt{n_{1}(k_1,0)}$ and $c_2(q,0)=c_3(q,0)=0$, where $n_1(k_1,0)$ is obtained by averaging the experimentally measured lattice momentum distributions of the initial spin-polarized samples. 

\begin{figure}
\includegraphics[width=8.4cm]{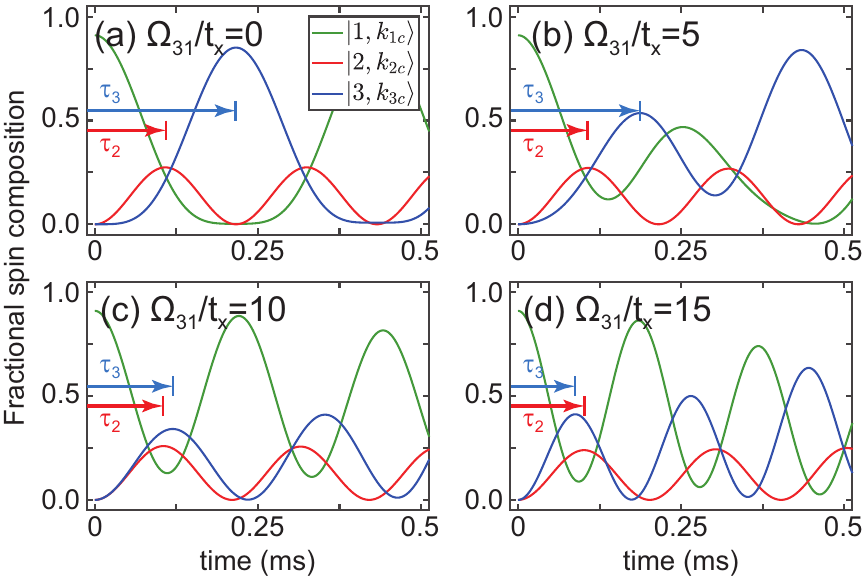}
\caption{ Calculated quench evolution of the fractional spin composition at $q_c=\pm\pi$ for various values of $\Omega_{31}/t_x$ in the synthetic three-leg Hall tube system. $t_x=2\pi\times264~$Hz and $\Omega_{12}=12.3 t_x$ as in the experiment present in Fig.~4. $\tau_s$ is the time when the fractional population in the spin $s$ state reaches its first maximum.\label{fig:simulation_tau}}
\end{figure}

Figure~S3 displays the numerical results of the quench dynamics for the various boundary conditions of the experiment. We observe that spin oscillations show damping in the three-leg Hall tube and three-leg open ladder cases [Figs.~\ref{fig:simulation_spin}(a) and \ref{fig:simulation_spin}(e)], whereas those in the two-leg open ladder case are not damped [Fig.~\ref{fig:simulation_spin}(c)]. We find that the effective damping originates from $\xi_2$ being nonzero in the numerical simulations. In the experiment, we also observe that damping is enhanced in the synthetic Hall tube and open-three leg ladder cases. In the calculations of $\langle k\rangle$$(t)$ and $\mathcal{C}(t)$ for the three-leg cases in Figs.~2(d) and Fig.~3(f), we include the damping effect phenomenologically as 
\begin{equation} \label{eq:kevol}
{g}_{e}(t)=\big( g(t)-\bar{g} \big) e^{-t/\tau_d} + \bar{g},
\end{equation}
where $g$ is $\langle k\rangle$ or $\mathcal{C}$ directly obtained from the numerical simulation and $\bar{g}$ is the mean value determined from the experiment. We find that $\tau_d=0.15$~ms for the synthetic three-leg Hall tube and $\tau_d=0.3$~ms for the open three-leg ladder show reasonable agreement with the experimental data.

Figure~\ref{fig:simulation_tau} displays the numerical results of the quench dynamics at $q_c=\pm\pi$, i.e., $\{k_{1c},k_{2c}, k_{3c}\}=\{1/3,\pm 1,-1/3\}k_\mathrm{L}$ for various values of $\Omega_{31}$. The time scales of spin oscillations are characterized with $\tau_s$ at which spin population in $|s\rangle$ reaches its first maximum.  At the critical point $\Omega_{31}=\Omega_-$ of the topological phase transition, $\tau_{2}=\tau_{3}$ is observed, which is a consequence of the associated band gap closing.

\end{document}